\documentclass[12pt]{article}
\usepackage{epsfig}

\begin{document}

\title{On Construction of ICRF-2}
\author{Zinovy Malkin}
\date{Pulkovo Observatory, St. Petersburg, Russia}
\maketitle

\begin{abstract}
In this paper, several issues are considered, related to the construction
of the next ICRF generation, ICRF-2. Between them, the following points
are touched: ICRF-2 structure, ICRF Core sources selection, and some expected
user's requirements.
\end{abstract}

\vfill
\noindent \hrule width 0.4\textwidth
~\vskip 0.2ex
\noindent {\small 5th IVS General Meeting, St.~Petersburg, Russia, 3--6 March 2007}
\eject

\section{Introduction}
By now, more than 6 million geodetic and astrometric VLBI observations were made,
much more than were available for the construction of ICRF, and methods of data analysis
were substantially advanced. This opens up new opportunities for ICRF improvement.
For this purpose, two ad hoc Working Groups were organized by IAU, and jointly IERS
and IVS\footnote{http://rorf.usno.navy.mil/ICRF2/}, and preparation to ICRF-2
creation was started \cite{Ma08}. In particular, one of the main tasks of the preparation
campaign is to investigate possible strategies that are or may be used for ICRF-2,
the second ICRF realization. In this paper several issues related to construction
of ICRF-2 are touched.

\section{ICRF-2 Structure}

It is suggested that ICRF-2 comprises two source lists.
%, which follows the structure of the latest optical Fundamental Catalogs.
The first list may be called {\it ICRF Core}.
%, which would have about the same status as the present ICRF {\it defining} source list.
The main features of the ICRF Core are:
\begin{itemize}
\itemsep=-0.5ex
\item realizes ICRS definitions;
\item consists of about 400 sources  evenly distributed over the sky,
   i.e. 1 source per 100~deg$^2$;
\item provides the long-time system stability.
\end{itemize}

The second list may be called {\it ICRF Extension} or {\it ICRF Supplement}.
The main features of the ICRF Extension are:
\begin{itemize}
\itemsep=-0.5ex
\item consists of about 3600 sources;
\item keeps ICRF system;
\item provides ICRF densification for extended user's needs;
\item serves as a set of calibrators for the phase-reference VLBI;
\item may have more dense source distribution in the ecliptic
   belt to be used in the space navigation.
\end{itemize}

These two lists give 4000 sources in total, i.e. 1 source per 10~deg$^2$.
Suggested total number of sources is close to the number of sources already
observed in the IVS and VCS sessions, and for most of them reliable positions
are derived, mostly for declination $> -40$~deg.

It seems to be reasonable to schedule in 2008 several dedicated IVS sessions
to obtain positions for 300--500 sources with reasonable precision in the southern
hemisphere zone with declination $< -40$~deg to complete ICRF Supplement list
in the sky regions poorly filled with astrometric radio sources.

In principle, it is advisable to have as many Core sources as possible
in order to minimize the impact on the individual source instability.
At the moment, 400 sources may be a good compromise between this requirement
and the real number of the well observed sources.
As the VLBI system becomes more sensitive, which allows us to observe more weak
compact sources, and number of well observed and investigated sources grows
new sources should be included in the ICRF Core catalog to provide better system
stability.
However, it is important to add new sources to the ICRF Core in such a way
to keep uniform distribution of the Core sources over the sky.

\section{Core Source Selection}

A key issue is how to select ICRF-2 Core sources.
Different criteria for source selection have been proposed by many authors, based on
\begin{enumerate}
\itemsep=-0.5ex
\item time series analysis: velocity of the apparent source motion, scatter parameters (indices)
   such as (w)rms, $\chi^2$ and Allan deviation; other statistics;
\item observation history: time span of observations, number of sessions or observations
   (correlation between them is at a level of 0.9), observation density, etc.;
\item physical characteristics: structure index, redshift;
\item other criteria, e.g. position uncertainty from global analysis.
\end{enumerate}

In this paper, only criteria of the first two groups are considered.
These criteria were applied to the time
series submitted by IVS Analysis Centers to the ICRF-2 data pool.
For all the computations, original source positions derived from
analysis of single session and reported by the analysis centers were used.
Reported uncertainties in the source positions were used for weighting.

For the criteria of the first group, when possible, 2D estimates were used,
e.g. source velocity was computed as $V=\sqrt{(V_\alpha\cos\delta)^2+V_\delta^2}$,
WRMS of source position computed analogously, Allan deviation computed
using the 2D weighted estimate WMADEV as developed in \cite{Malkin08}.
The uncertainty in the velocity estimate was also used as a new scatter index.

\begin{figure}
\centering
\epsfclipon \epsfxsize=0.85\textwidth \epsffile{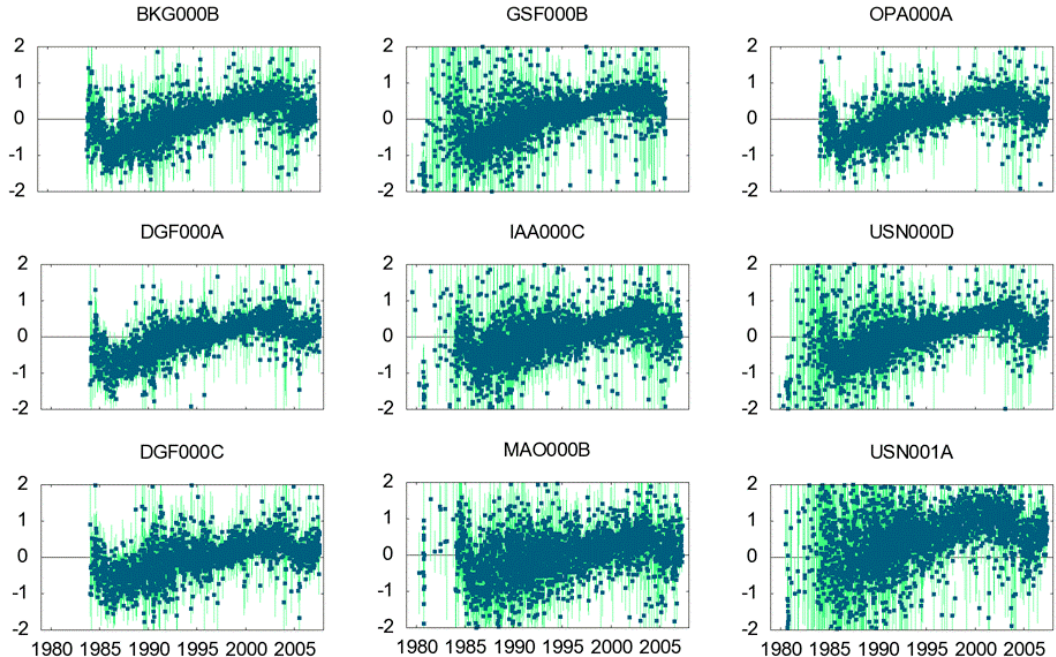}
\caption{Several time series for 0923+392 (3C39.25) RA.
In this example, several solutions show similar behavior of the source coordinate
but different scatter.}
\label{fig:0923+392}
\end{figure}

\begin{figure}
\centering
\epsfclipon \epsfxsize=0.85\textwidth \epsffile{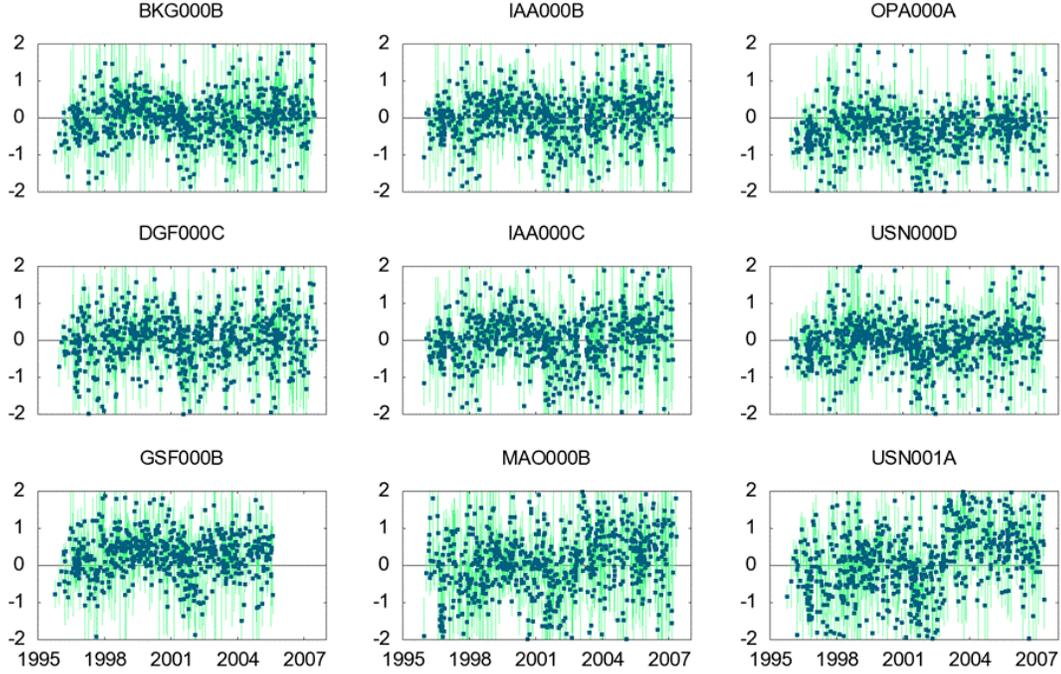}
\caption{Several time series for 0003-066 DE.
In this example, several solutions show both different behavior of the source coordinate
and scatter.}
\label{fig:0003-066}
\end{figure}

The first conclusion made from this analysis is that source behavior and consequently
source position variation (scatter) indices may differ significantly between the analysis
centers, as can be seen from Figs.~\ref{fig:0923+392} and~\ref{fig:0003-066}.


In Table~\ref{tab:0656+082_vel} an example is given for velocity estimate using several time series.

\begin{table}
\centering
\caption{2D velocity of the source 0656+082 computed from different source position time series.}
\label{tab:0656+082_vel}
\tabcolsep=10pt
\begin{tabular}{|l|c|c|}
\hline
Series  & Nepochs & V, $\mu$as/yr \\
\hline
mao000b    & 197  & ~45 $\pm$ 38 \\
usn000d    & 189  & ~53 $\pm$ 28 \\
dgf000a-g  & 188  & ~80 $\pm$ 30 \\
opa001a    & 186  & 101 $\pm$ 29 \\
iaa000c    & 168  & 107 $\pm$ 24 \\
usn001a    & 189  & 124 $\pm$ 43 \\
iaa000b    & 168  & 163 $\pm$ 24 \\
aus001a    & 179  & 165 $\pm$ 42 \\
gsf000b    & 131  & 235 $\pm$ 50 \\
\hline
\end{tabular}

\end{table}

As can be seen from Table~\ref{tab:0656+082_vel}, despite large number of epochs
and considerable time span about 7.5 yr, velocity estimates for source 0656+082
differ by several times, which shows that analysis strategy has large impact
even on the estimated radio source velocity, a parameter, which is often attributed
to physical processes such as jet motion, and which should seem most robust with
respect to the analysis strategy. Large difference between available time series
can be observed also for other scatter indices, cf. also Figs. \ref{fig:0923+392}
and \ref{fig:0003-066}.

So, results of source position time series analysis heavily depend on method
used for time series computation and should be used with care.
Under this circumstance, maybe the number of sessions, network diversity and time span
should be a first criteria to select ICRF Core candidates.
Only sufficient time span and number of sessions can provide reliable conclusion
on the source position variations.
This inference can also be drawn from the position time series for some sources
where one can see quite different velocity and scatter at different part of
the interval of observations.

\begin{figure}[h!]
\centering
\epsfclipon \epsfxsize=0.8\textwidth \epsfysize=0.6\epsfxsize \epsffile{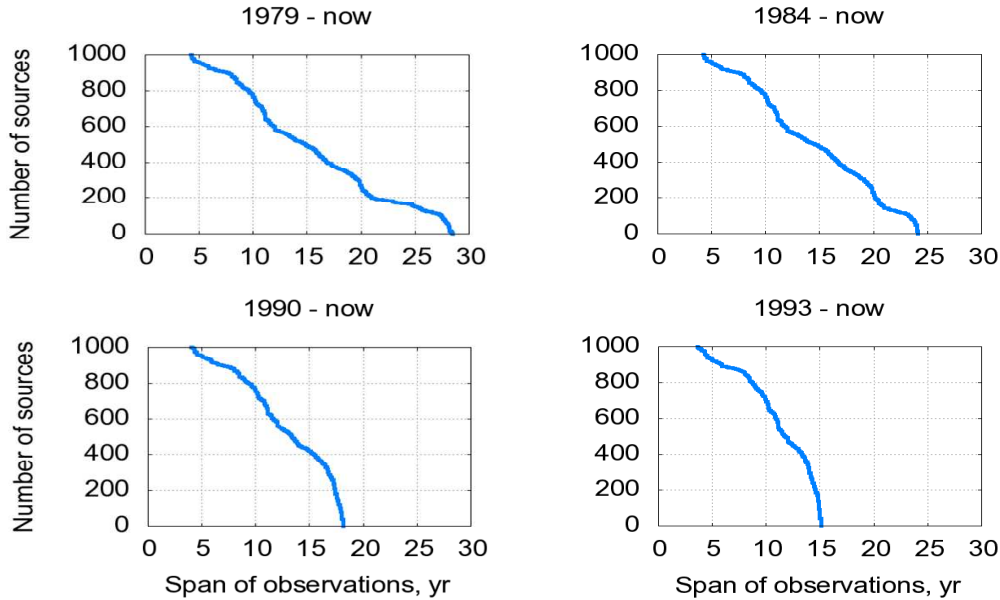}
\caption{Number of sources for given start epoch and data span.
   Four plots are shown for different criteria for the data selection discussed in literature.}
\label{fig:N_sou}
\end{figure}

At the first glance we can find sufficient number of sources with large
observations history, see e.g. Fig.~\ref{fig:N_sou}, but only at the first glance.
The well known problem is that distribution of the well observed sources are far
from uniform, with clear deficiency in the southern hemisphere.

Evidently, at the moment we have not enough sources of the highest quality
(compactness and position error) to fill in all the evenly distributed
100-deg$^2$ sky cells.
The following strategy for the nearest years can be considered:
\begin{itemize}
\itemsep=-0.5ex
\item for the first ICRF-2 version, one, the best of available,
source in each 100-deg$^2$ cell should be taken; at this stage a list of
sources of insufficient quality (time span, number of sessions, position error)
is identified;
\item at the next stage, the sources of insufficient quality should be
observed in 2009-2010 to be able to issue the second version of ICRF-2
by the end of 2010--beginning of 2011.
\end{itemize}

This should be mentioned that if two or more good sources fell
to the same cell, they will not be lost for users since they will
be included in the ICRF Extension.
The general logic of this approach is to give priority to
even source distribution, which presumably has advantage in
maintenance of the ICRF orientation and further comparison with
GAIA CRF realization.

\section{ICRF for Users}

From user's point of view, a source position catalog is a tool
to predict the source coordinates at given epoch with known error.
Currently, a user can use the source position error given in the catalog.
These errors are usually obtained from least square adjustment,
and can be considered as the precision only.
To assess the accuracy of the source coordinates, one can analyze
the source coordinates time series. For example, the uncertainty in the
position of the source 0923+392 (3C39.25) given in the ICRF-Ext.2 catalog is~0.035 mas.
However, one can see from Fig.~\ref{fig:0923+392} that this value is
well underestimated, and actual position error for certain
epochs may be as large as 1~mas.
The error in source position based on time series analysis
may be computed by the method proposed in~\cite{Malkin01}, which
allows us to account for both precision and scatter of the session
position estimates.

Based on this considerations we can conclude that the End User Error (EUE) concept,
based on the realistic estimate of the source position accuracy,
may be more adequate for ICRF instead of or in addition to the {\it defining}
or {\it stable} concepts, since it show just what a user gets using the catalog.
Also, we can consider the Index of Position Variability (IPV) based on existing
and/or new methods of analysis and computed as a continuous function, instead of
the stepwise index 1-2-3-... widely used now, which should be given in a ICRF catalog
to give a user a quantitative measure, or at least an impression, of the real error
in the source position.

\end{document}